\documentclass[12pt]{article}
\usepackage{titling}
\usepackage{amsmath}
\usepackage{slashed}
\usepackage{amssymb}
\usepackage{epsfig}
\usepackage{graphicx}
\usepackage{multirow}
\usepackage{color}
\usepackage[normal]{subfigure}
\usepackage{rotating}
\usepackage{hyperref}
\usepackage[margin=0.9in]{geometry}
\usepackage[table]{xcolor}
\usepackage{enumitem}
\usepackage[utf8x]{inputenc}
\usepackage[compress,numbers,sort]{natbib}
\usepackage{authblk}
\usepackage{colortbl}
\usepackage{pdflscape}
\usepackage{color}

\definecolor{nicered}{rgb}{0.6,0.1,0.1}
\definecolor{nicegreen}{rgb}{0.1,0.5,0.1}
\definecolor{mediumcandyapplered}{rgb}{0.99, 0.12, 0.07}
\definecolor{red}{rgb}{1.0, 0, 0}
\hypersetup{colorlinks,citecolor= nicegreen,linkcolor= nicered}

\def\eq#1{{Eq.~(\ref{#1})}}
\def\eqs#1#2{{Eqs.~(\ref{#1})--(\ref{#2})}}

\def\Table#1{{Table~\ref{#1}}}
\def\Tables#1#2{{Tables~\ref{#1}--\ref{#2}}}
\def\sect#1{{Sect.~\ref{#1}}}
\def\sects#1#2{{Sects.~\ref{#1}--\ref{#2}}}
\def\app#1{{Appendix~\ref{#1}}}

\def\abs#1{\left| #1\right|}

\def\Im{\mbox{Im}\,}
\def\Re{\mbox{Re}\,}
\def\Tr{\mbox{Tr}\,}

\renewcommand{\bar}{\overline}

\definecolor{LightCyan}{rgb}{0.88,1,1}
\definecolor{piggypink}{rgb}{0.99, 0.87, 0.9}
\definecolor{applegreen}{rgb}{0.55, 0.71, 0.0}
\definecolor{darkpastelgreen}{rgb}{0.01, 0.75, 0.24}
\definecolor{green-yellow}{rgb}{0.68, 1.0, 0.18}

\newcommand{\beq}{\begin{equation}}
\newcommand{\eeq}{\end{equation}}
\newcommand{\bea}{\begin{eqnarray}}
\newcommand{\eea}{\end{eqnarray}}

\title{\bf{What is the scale of new physics behind \\ the $B$-flavour anomalies?}}

\author[1]
{Luca Di Luzio\thanks{luca.di-luzio@durham.ac.uk}}
\author[2]
{Marco Nardecchia\thanks{marco.nardecchia@cern.ch}}

\affil[1]{\emph{\normalsize Institute for Particle Physics Phenomenology, Department of Physics, 
Durham University, DH1 3LE, Durham, United Kingdom}}
\affil[2]{\emph{\normalsize Theoretical Physics Department, CERN, Geneva, Switzerland}}

\date{}

\begin{document}

\maketitle
\begin{abstract}
\normalsize
Motivated by the recent hints of lepton flavour non-universality in $B$-meson 
semi-leptonic decays, we study the constraints of perturbative unitarity on the new physics 
interpretation of the anomalies in $b \to c \ell \bar \nu$ and $b \to s \ell \bar \ell$ transitions. 
Within an effective field theory approach we find that $2 \to 2$ fermion scattering amplitudes 
saturate the unitarity bound below $9$ TeV and $80$ TeV,  
respectively for $b \to c \ell \bar \nu$ and $b \to s \ell \bar \ell$ transitions. 
Stronger bounds, up to few TeV, 
are obtained when the leading effective operators are oriented in 
the direction of the third generation, as suggested by flavour models.
We finally address unitarity constraints on simplified models explaining the anomalies 
and show that the new physics interpretation is ruled out in a class of perturbative realizations. 
\end{abstract}

\clearpage

\tableofcontents

\clearpage

\section{Introduction}

In the recent years we have witnessed a growing pattern of experimental anomalies in flavour physics, 
which can be schematically summarized as follows: 
\begin{enumerate}
\item Semi-leptonic $B$-decays in flavour changing neutral currents (FCNC) $b \to s \ell \bar \ell$, 
suggesting a deficit of muons compared to electrons. The main observables are: $i)$ the angular distributions of 
$B \to K^* \mu \bar \mu$ \cite{Aaij:2014pli,Aaij:2013qta,Aaij:2015oid}, $ii)$ the rate of semi-leptonic decays such as 
$B \to K^* \mu \bar \mu$ \cite{Aaij:2014pli} and $B_s \to \phi \mu \bar \mu$ \cite{Aaij:2015esa} and $iii)$ 
the lepton flavour universality (LFU) violating observables $R_K$ \cite{Aaij:2014ora} and $R_{K^*}$ \cite{Aaij:2017vbb}, 
which are defined by the ratios $\mathcal{B}(B \to K^{(*)} \mu \bar \mu) / \mathcal{B}(B \to K^{(*)} e \bar e)$. 
We remark that the Standard Model (SM) theoretical uncertainty for $R_{K^{(*)}}$ is very small 
(few percent due to QED radiative corrections \cite{Bordone:2016gaq}). Updated fits based on effective field theory (EFT) 
analyses, including the most recent $R_K{^*}$ measurement, can be found in \cite{Capdevila:2017bsm,Altmannshofer:2017yso,DAmico:2017mtc,Ciuchini:2017mik,Geng:2017svp}. 
\item Semi-leptonic $B$-decays in flavour changing charged currents (FCCC) $b \to c \ell \bar \nu_{\ell} $, 
suggesting an excess of taus compared to muons and electrons. The main observables are the LFU violating 
ratios $R_{D^{(*)}}$ \cite{Lees:2013uzd,Aaij:2015yra,Hirose:2016wfn}, defined as 
$\mathcal{B}(B \to D^{(*)} \tau \bar \nu) / \mathcal{B}(B \to D^{(*)} \ell \bar \nu)$, 
with $\ell = e, \mu$. In this case it is non-trivial that three different experiments agree well among each other.  
A recent EFT fit of $R_{D^{(*)}}$ can be found for instance in \cite{Bernlochner:2017jka}.
Very recently, there has also been a new measurement of $R_{D^*}$ by the LHCb collaboration \cite{LHCb:CERN}, 
which is remarkably compatible with the previous ones.  
However, this single measurement does not affect much the global fit.  
\end{enumerate}
In both cases the statistical significance reaches the $4\sigma$ level, while 
from a theoretical standpoint it is very intriguing that both the set of anomalies can be interpreted within 
a coherent framework. In particular, one can envisage two orthogonal structures: 
$i)$ a vertical (gauge) one: global fits seem to prefer effective operators 
featuring only $SU(2)_L$ doublets and $ii)$ a horizontal (flavour) one: 
data hints to violation of LFU with a similar hierarchical pattern as in the SM, with new physics 
contributions negligible in electrons (basically no effects), sizeable in muons (observable only in $b \to s \mu \bar \mu$) 
and large in taus (effects in $b \to c \tau \bar \nu_\tau$ and potentially in $b \to s \nu_\tau \bar \nu_\tau$).  
These facts motivated the community to speculate about the simultaneous explanation of these two sets of anomalies 
and their connection with the origin of the SM flavour. It is then maybe not too early to dream about new physics 
and ask ``what is the scale of new physics behind the $B$-flavour anomalies?'' 

Here, we address 
this question by using an old tool of theoretical physics, namely perturbative unitarity. 
Perhaps most famously, constraints imposed by perturbative unitarity in $WW$ scattering have been used in the past to infer 
an upper bound on the Higgs boson mass or, alternatively, on the scale where the SM description of weak interactions 
needed to be completed in the ultraviolet (UV) in terms of some new strongly coupled dynamics \cite{Lee:1977yc,Lee:1977eg}.
What we are going to consider here instead resembles in some sense the Fermi theory of weak interactions \cite{Fermi:1934sk}. 
In fact, already in the 1930's, 
from the low-energy measurement of $G_F$ one could have inferred what was 
the scale of ``new physics'' behind the Fermi theory. 
By looking 
at $2 \to 2$ scatterings via four-fermion effective operators in the Fermi theory one finds 
that unitarity is violated\footnote{We will sometimes improperly use the term ``unitarity violation'', 
by which we mean perturbative unitarity (cf.~the discussion in \sect{UBs}).} 
at energies of the order of $\Lambda_U = 900$ GeV (see e.g.~\cite{Becchi:2006sk}). 
As is well known, the dynamical degrees of freedom of the SM turned out to be weakly coupled 
and hence much lighter than the unitarity bound, e.g.~$M_W \ll \Lambda_U$. 

In this paper, we do something similar to the unitarity analysis in the Fermi theory 
by considering the four-fermion operators of the $d=6$ 
SM-invariant EFT (SMEFT) semi-leptonic basis, 
under the hypothesis of a short-distance new physics explanation of the experimental anomalies in 
semi-leptonic $B$-meson decays.\footnote{Unitarity bounds for the EFT intrerpretaton of $b \to s \ell \bar \ell$ 
anomalies were briefly mentioned in Ref.~\cite{Altmannshofer:2017yso}.}  
Note that the analysis can be independently carried out for the 
$b \to s \ell \bar \ell$ and $b \to c \ell \bar \nu$ anomalies, 
and we do not necessarily rely on a common explanation of the two. 
In short, once the Wilson coefficient of an effective operator is fixed by the fit to the anomaly, 
we can use it in order to extract the scale of unitarity violation 
without the need of passing through the ambiguous separation of mass vs.~coupling. 
The common lore is that on-shell new degrees of freedom should appear below the scale of unitarity violation 
(see however \cite{Aydemir:2012nz} for exceptions), 
with interesting consequences for direct searches at LHC and future colliders. 

A simple message that we would like to emphasize is that scattering amplitudes employing SM 
invariant effective operators lead to scales of unitarity violation $\Lambda_U$
which are typically smaller than the naive dimensional analysis (NDA) 
estimate of the strong coupling regime $g_\star = 4 \pi$, i.e.~$M_\star = 4 \pi \Lambda_{\mathcal{O}}$, 
where $\Lambda_{\mathcal{O}}$ 
denotes the scale of the SM invariant effective operator required to fit the anomaly normalized to unit Wilson coefficient. 
This enhancement is also in part due to the correlation of the scattering amplitudes in the gauge group space, 
which is important to take into account when thinking about the energy reach of LHC or future colliders.
A related point is the flavour structure of the effective operators. If these are oriented along the third generation fermion families 
(as motivated in various flavour models), one typically predicts a strong enhancement of the unitarity bound 
which can even reach few TeV (cf.~\Table{tablescales}). 

Similarly to the EFT analysis, unitarity arguments can also be used in order to set perturbativity 
constraints on the parameter space of simplified models explaining the flavour anomalies. 
Note, however, that in the latter case the scattering amplitudes 
do not grow with the energy but reach asymptotic values proportional to the Yukawa-like couplings 
of the new mediators. It is possible then to translate the unitarity bounds on the coupling into an upper bound 
on the mass of the new states (once the ratio coupling/mass is fixed in terms of the fit to the relevant anomaly). 
Remarkably, in some cases the upper bound on the new mediators' mass is so strong that the perturbative interpretation of 
the anomaly within a given simplified model can be ruled out, or soon tested at the LHC. 

The layout of the paper is the following: in \sect{tale} we start by introducing and comparing 
different kind of scales in the EFT. After discussing in \sect{FS} 
motivated flavour structures for the effective operators, we briefly introduce the partial-wave-unitarity 
tool in \sect{UBs}. We continue in \sects{semilepSMEFT}{simplmodels} where we derive the unitarity bounds 
respectively in the EFT and for simplified models addressing the $B$-flavour anomalies. 
We finally conclude in \sect{concl}, 
where we also provide a summary of our results. In \app{tripop}, as a paradigmatic example, 
we report the details of the unitarity bound calculation in the presence of an $SU(2)_L$ triplet effective operator.

\section{A tale of scales}
\label{tale}

In what follows we will focus for simplicity on purely left-handed operators, since they provide the best fit for 
both the anomalies in $b \to s \mu \bar \mu$ and $b \to c \tau \bar \nu$ transitions. The analysis can be easily 
generalized to scenarios including more operators by using the results given in \sect{semilepSMEFT}. 
In order to start the discussion it is useful to identify and compare four (conceptually different) scales 
in the EFT:\footnote{Some of the results presented here will be derived in the following sections.} 
\begin{enumerate}
\item $\Lambda_{A}$: the ``Fermi constant'' of the process. \\
This is the scale required to explain the anomaly, to be evaluated at the typical energy of the process which is fixed 
by the $B$-meson mass. The low-energy EFT description is based on $SU(3)_C \times U(1)_{EM}$ invariant 
operators. The index $A$ on $\Lambda_A$ runs over the anomalies, schematically $A=\{ R_{D^{(*)}},R_{K^{(*)}} \}$, and the EFT 
Lagrangian featuring purely left-handed operators reads
\beq
\label{LeffRDRK}
\mathcal{L}_{\rm eff} \supset 
- \frac{1}{\Lambda^2_{R_{D^{(*)}}}} 2 \, \overline{c}_L \gamma^{\mu} b_L \overline{\tau}_L \gamma_{\mu} \nu_L 
+ \frac{1}{\Lambda^2_{R_{K^{(*)}}}}\overline{s}_L \gamma^{\mu} b_L \overline{\mu}_L \gamma_{\mu} \mu_L 
+ \textrm{h.c.} \, ,   
\eeq
where we assumed alignment with the phases of the CKM elements that appear in the corresponding SM operators.
Note that the fit of the $R_{D^{(*)}}$ and $R_{K^{(*)}}$ anomalies requires an opposite sign 
interference with the SM contribution. We also included an extra factor of $2$ in the definition of the charged-current operator, 
so that the latter has the same normalization of the neutral-current operator when considering a SMEFT.  
The best fit values of the $R_{D^{(*)}}$ \cite{Alonso:2015sja} and $R_{K^{(*)}}$ \cite{DAmico:2017mtc} anomalies yield respectively
\begin{align}
\Lambda_{R_{D^{(*)}}} &= 3.4 \pm 0.4 \textrm{ TeV} \, , \\
\Lambda_{R_{K^{(*)}}} &= 31 \pm 4  \textrm{ TeV} \, ,
\end{align}
where the errors are at 1$\sigma$. In the following we will only consider central values. 

\item $\Lambda_{\mathcal{O}}$: the scale of the SMEFT operator. \\
This is the scale required to explain the anomaly using an EFT at higher 
energies\footnote{QCD running effects on the Wilson coefficients are of the order of $1 +\frac{\alpha_s}{4 \pi} \times \log \frac{\Lambda_{\mathcal{O}}}{m_b}$. 
For $\Lambda_{\mathcal{O}}= 1$ TeV, this corresponds to an $\mathcal{O}(5\%)$ correction that will be neglected in the 
following.} ($SU(3)_C \times SU(2)_L \times U(1)_Y$ invariant), with Wilson coefficient normalized to one.       
The index $\mathcal{O}$ on $\Lambda_{\mathcal{O}}$ is associated with an operator of the 
SMEFT semi-leptonic basis and 
runs over all the possible Lorentz and flavour structures. 
For definiteness we will consider here an $SU(2)_L$ triplet operator ($Q$ and $L$ denoting $SU(2)_L$ doublets)
\beq 
\label{opQijL33}
\mathcal{L}_{\rm SMEFT} \supset \frac{1}{\Lambda^2_{Q_{ij} L_{kl}}} \left( \bar Q_i \gamma^\mu \sigma^A Q_j \right) 
\left( \bar L_k \gamma_\mu \sigma^A L_l \right) + \textrm{h.c.} \, , 
\eeq 
and two reference flavour structures such that the operator is aligned in the direction 
of the flavour eigenstates responsible for the anomalies, namely 
$\mathcal{O} = Q_{23} L_{33}$ (for $b \to c \tau \bar \nu$ transitions)
and $\mathcal{O} = Q_{23} L_{22}$ (for $b \to s \mu \bar \mu$ transitions).
The matching with \eq{LeffRDRK} yields 
\begin{align}
\label{LamQ23L33}
\abs{\Lambda_{Q_{23} L_{33}}} &= \Lambda_{R_{D^{(*)}}}  = 3.4 \textrm{ TeV} \, , \\
\label{LamQ23L33}
\abs{\Lambda_{Q_{23} L_{22}}} &= \Lambda_{R_{K^{(*)}}}  = 31 \textrm{ TeV} \, . 
\end{align}
As we will discuss in detail in \sect{FS}, depending on the specific flavour ansatz, the scale $\Lambda_{\mathcal{O}}$ 
can be effectively reduced with respect to the ``Fermi constant'' of the process. 
For example, the transition $b \to c \tau \bar \nu$ could originate from the operator $\mathcal{O} = Q_{33} L_{33}$, 
where the $3 \to 2$ transition in the up sector is due to a CKM mixing (in the basis where 
$Q_i = (V^\dag_{ij} u^j_L, d^i_L)^T$), which yields $\Lambda_{Q_{33} L_{33}} / \sqrt{\abs{V_{cb}}} = \Lambda_{R_{D^{(*)}}}$.
 
\item $\Lambda_U$: the scale of unitarity violation. \\
This is the scale where the EFT description breaks down.  
The important point is that it can be 
expressed in terms of the scale $\Lambda_{\mathcal{O}}$, 
without passing through the ambiguous separation between coupling and mass.  
Using the results of \sect{semilepSMEFT} (which are based on a non-trivial calculation of the scattering amplitude, including gauge group multiplicity 
factors) we obtain
\beq 
\Lambda_U  = \sqrt{\frac{4 \pi}{\sqrt{3}}} \abs{\Lambda_{Q_{ij} L_{kl}}} \, , 
\eeq
which yields 
\begin{align}
\label{LambdaU1}
\Lambda_U   &= 9.2 \textrm{ TeV} \qquad  (\mathcal{O} = Q_{23} L_{33} \ \ \text{case}) \, , \\
\label{LambdaU2}
\Lambda_U   &= 84 \textrm{ TeV} \qquad \ (\mathcal{O} = Q_{23} L_{22} \ \ \text{case}) \, .  
\end{align}
These are the most conservative bounds on the scale of new physics 
responsible for the anomalies in $ b\to c \tau \bar \nu $ and $ b \to s \mu \bar \mu$.
\item $M_\star$: the NDA mass scale in the strongly coupled regime. \\
This is the mass scale associated with the effective operator when saturating perturbativity. 
After reintroducing $\hbar$ in the NDA (see e.g.~\cite{Manohar:1983md,Cohen:1997rt,Panico:2015jxa}), one can formally 
distinguish among scales ($\Lambda$), masses ($M$) and couplings ($g$),
and set $M = g \Lambda$. 
By naively saturating perturbativity at $\abs{g_\star}= 4 \pi$, 
we can write
\beq
\frac{1}{\abs{\Lambda_{\mathcal{O}}}} =  \frac{4 \pi}{M_{\star}} \, ,
\eeq
which leads to 
\begin{align} 
\label{Mstar1}
M_\star &= 43 \textrm{ TeV} \qquad  \ (\mathcal{O} = Q_{23} L_{33} \ \ \text{case}) \, , \\
\label{Mstar2}
M_\star &= 390 \textrm{ TeV} \qquad  (\mathcal{O} = Q_{23} L_{22} \ \ \text{case}) \, .  
\end{align}
Note that $M_\star$ is a factor $5$ larger than the scale of unitarity violation in \eqs{LambdaU1}{LambdaU2}.
\end{enumerate}
Our results for the EFT analysis are summarized in \Table{tablescales} (cf.~also \sects{FS}{semilepSMEFT} for more details 
on the flavour structure of the effective operators and 
the unitarity bounds), where we report the values 
of the four different scales discussed above for the anomalies in either $b \to c \tau \bar \nu$ 
or $b \to s \mu \bar \mu$ 
transitions,
and depending on the flavour structure of the operator $\mathcal{O}$. The two main points to be observed are the following: 
$i)$ $\Lambda_U$ is sizably smaller than $M_\star$ and $ii)$ depending on the flavour structure of the operator $\mathcal{O}$, the scale $\Lambda_U$ 
approaches the energy reach of LHC. This motivates an interesting interplay of the flavour anomalies with direct searches, 
which is further explored in \sect{simplmodels} by employing simplified models.

\begin{table}[htp]
\begin{center}
\begin{tabular}{|c|c|c|c|c|c|c|c|}
\hline
Anomaly & $\mathcal{O}$ & FS$_{Q}$  & FS$_{L}$ & $\Lambda_A [\text{TeV}]$ &$\abs{\Lambda_{\mathcal{O}}} [\text{TeV}]$ & \cellcolor{piggypink} $\Lambda_U [\text{TeV}]$ & $M_\star [\text{TeV}]$ \\
\hline
\hline
$ b \to c \tau \bar \nu $ & $Q_{23}L_{33}$ & 1 & 1 & 3.4 & 3.4 & \cellcolor{piggypink} 9.2 & 43  \\
$ b \to c \tau \bar \nu $ & $Q_{33}L_{33}$ & $\abs{V_{cb}}$ & 1 & 3.4 & 0.7 & \cellcolor{piggypink} 1.9 & 8.7  \\
\hline
$ b \to s \mu \bar \mu $ & $Q_{23}L_{22}$ & 1 & 1 & 31 & 31 & \cellcolor{piggypink} 84 & 390  \\
$ b \to s \mu \bar \mu $ & $Q_{33}L_{22}$ & $\abs{V_{ts}}$ & 1 & 31 & 6.2 & \cellcolor{piggypink} 17 & 78  \\
$ b \to s \mu \bar \mu $ & $Q_{33}L_{33}$ & $\abs{V_{ts}}$ & $^\ddag m_{\mu}/m_{\tau}$& 31 & 1.5 & \cellcolor{piggypink} 4.1 & 19  \\
$ b \to s \mu \bar \mu $ & $Q_{33}L_{33}$ & $\abs{V_{ts}}$ & $^\ast (m_{\mu}/m_{\tau})^2$& 31 & 0.4 & \cellcolor{piggypink} 1.0 & 4.7  \\
\hline
\end{tabular}
\end{center}
\caption{Summary of the different new physics scales associated with the $B$-flavour anomalies in the EFT analysis: 
$\Lambda_A$ is the scale of the effective operator needed to fit the low-energy observable, 
$\Lambda_{\mathcal{O}}$ is that required by a SMEFT, $\Lambda_U$ is the scale of unitarity violation and 
$M_\star$ is the NDA mass scale of the operator in the strongly coupled regime. 
$\mathcal{O}$ denotes the flavour structure of the triplet operator in \eq{opQijL33}, while 
FS$_{Q}$ and FS$_{L}$ are flavour suppression factors in the quark and lepton sector which 
rescale the aligned entries (those corresponding to $\text{FS}_{Q,L}=1$) by a factor 
$\sqrt{\text{FS}_{Q} \times \text{FS}_{L}}$.
The cases marked by $\ddag$ and $\ast$ correspond respectively 
to the ansatz of left-right symmetric partial compositeness and minimal flavour violation in the charged lepton sector 
(see \sect{FS} for details).  
}
\label{tablescales}
\end{table}%

\section{On the flavour structure of the effective operators}
\label{FS}

The $R_{D^{(*)}}$ and $R_{K^{(*)}}$ anomalies can be interpreted via new physics 
contributions in quark flavour transitions involving the third and second generation, 
respectively $b \to c$ for FCCC and $b \to s$ for FCNC.
In models with motivated flavour structures, it is natural to expect sizable effects in channels not directly related to the flavour anomalies.
In particular, it may happen that operators involving fermions of the third family are enhanced compared to flavour violating ones. 
This implies that a stronger unitary bound can be derived from $2 \to 2$ scatterings of fermions of the third generation. 
For example, when considering the channel related to the anomaly in $b \to c \tau \bar{\nu}_{\tau}$ we always 
get a unitarity bound from the scattering $b \overline{c} \to \tau \bar{\nu}_{\tau}$, but we can reasonably expect that scatterings of the form $b \overline{b} \to \tau \bar\tau$ give stronger unitarity constraints.
In order to create a link between the different channels, a flavour structure has to be assumed. 
In the following, we review some well-known frameworks:

\begin{enumerate}
\item Minimal Flavour Violation (MFV)

The MFV hypothesis \cite{DAmbrosio:2002vsn} states that the strength of new physics effects are linked to the SM Yukawa couplings, 
which act as sources of breaking of the enlarged symmetry of the gauge-kinetic terms for fermions, $SU(3)^3$ for quarks. 
In particular, for quark doublets we get that flavour violating interactions are generated at the leading order (in powers of Yukawas) by
\beq
\overline{Q}_i \left( a \, Y_U Y^{\dagger}_U + b \, Y_D Y^{\dagger}_D \right)_{ij} Q_j \, ,
\eeq
where we omitted $SU(2)_L$ and Lorentz indices. Here, $a$ and $b$ are coefficients of similar size.
This implies a suppression of flavour violating quark currents compared to flavour conserving ones 
\beq
\frac{\overline{c}_L \gamma^{\mu} b_L}{\overline{t}_L \gamma^{\mu} b_L} \sim \frac{V_{cb}}{V_{tb}} \simeq V_{cb} 
\, , \qquad  
\frac{\overline{s}_L \gamma^{\mu} b_L}{\overline{b}_L \gamma^{\mu} b_L} \sim \frac{V^*_{ts}}{V^*_{tb}} \simeq V^*_{ts} \, .
\eeq

\item $SU(2)_Q$ flavour symmetry

In the limit of vanishing SM Yukawas for the first two quark generations, an $SU(2)^3$ global symmetry is restored. 
This approximate symmetry (or a subgroup of it) might be promoted to be a fundamental symmetry in the UV. 
In particular, there might be an $SU(2)_Q$ symmetry that distinguishes the quark doublets of the first two generations from the third one, 
and which has to be eventually broken in order to reproduce the observed pattern of SM masses and mixings. 
If the breaking is achieved via a spurion field $\vec{X}$ that transforms as the fundamental representation of $SU(2)_Q$, 
we get (see e.g.~\cite{Barbieri:2011ci}) that the the typical size of $|\vec{X}|$ is of $\mathcal{O}(\lambda^2)$, 
where $\lambda \sim 0.2$ is the Cabibbo angle. 
In this case, we also expect that new physics effects in FCCC and FCNC scale like
\beq
\label{U2scaling}
\frac{\overline{c}_L \gamma^{\mu} b_L}{\overline{t}_L \gamma^{\mu} b_L} \sim \lambda^2 
\, , \qquad  
\frac{\overline{s}_L \gamma^{\mu} b_L}{\overline{b}_L \gamma^{\mu} b_L} \sim \lambda^2 \, .
\eeq

\item Partial compositeness (PC)

A dynamical explanation of the flavour structure of the SM is provided by the paradigm of PC \cite{Kaplan:1991dc} 
in the context of composite Higgs models. 
In this framework the SM fields are linear combinations of elementary and composite states. The admixture elementary-composite 
of every SM state is regulated by a parameter $\epsilon^A_i$, where $A$ runs over the SM fermion fields ($A=Q,L,u,d,e$) and $i$ is a family index.
In terms of the mixing parameters, the Yukawas of the SM are given by
$(Y_U)_{ij} \sim \epsilon^Q_i \epsilon^{u}_j$ and $(Y_D)_{ij} \sim \epsilon^Q_i \epsilon^{d}_j$.
It is possible to show (see e.g.~\cite{KerenZur:2012fr}) 
that the $\epsilon^Q_i$ are linked to the size of the CKM matrix elements,  
i.e.~$\epsilon_2 / \epsilon_3 \sim \lambda^2$ and $\epsilon_1 / \epsilon_3 \sim \lambda^3$.
New physics effects are hence related to the size of the $\epsilon^A_i$ coefficients, 
and for quark left-handed currents one expects 
a similar scaling for FCCC and FCNC as in \eq{U2scaling}. 
\end{enumerate}
We conclude that for all the three frameworks above the transition between the third and second generation 
is suppressed by a factor $\mathcal{O}(\lambda^2)$ compared to the diagonal case involving only the third family. 
This implies that stronger unitarity bound can be derived from $2 \to 2$ scattering of the third family. For the presentation of our results in \Table{tablescales} we fix the numerical values to the MFV case, 
leading to a $\abs{V_{cb}}$ suppression in FCCC and a $\abs{V_{ts}}$ one in FCNC.

On the other hand, the situation in the lepton sector crucially depends on 
the unknown origin of neutrino masses. 
Note that the new physics effects required by the $B$-flavour anomalies do 
not violate the accidental $U(1)_{e,\, \mu,\, \tau}$ 
symmetry of the SM which arises in the $m_\nu \to 0$ limit (or, equivalently, in the 
decoupling limit of lepton-number-violating effective operators). 
It is hence reasonable to assume that the source of LFU breaking required by the 
$B$-flavour anomalies is connected to the charged lepton masses. 
Two structures can be easily motivated: 
\beq
\frac{\bar \ell^i_L \gamma^\mu \ell^i_L}{\bar \ell^j_L \gamma^\mu \ell^j_L} 
\sim \frac{(\epsilon^L_i)^2}{(\epsilon^L_j)^2} 
\sim \frac{m_{\ell_i}}{m_{\ell_j}} 
\qquad \text{or} \qquad 
\frac{\bar \ell^i_L \gamma^\mu \ell^i_L}{\bar \ell^j_L \gamma^\mu \ell^j_L} 
\sim \frac{(Y_E Y_E^\dag)_{ii}}{(Y_E Y_E^\dag)_{jj}}
\sim \left(\frac{m_{\ell_i}}{m_{\ell_j}}\right)^2 \, , 
\eeq
where the first option corresponds to PC with $\epsilon^L_i \sim \epsilon^e_i$
(implying $(Y_E)_{ij} \sim \epsilon^L_i \epsilon^L_j$) 
and the second one to MFV in the charged lepton sector. 
In \Table{tablescales} we use these two benchmarks, though different patterns can be of course envisaged.

\section{Partial wave unitarity}
\label{UBs}

Here we briefly recap the partial-wave unitarity formalism. More details can be found e.g.~in \cite{Itzykson:1980rh,Chanowitz:1978mv}.  
Let us denote by $\mathcal{M}_{fi}(\sqrt{s},\cos\theta)$ the matrix element of a $2 \to 2$ scattering amplitude in 
momentum space, where $\sqrt{s}$ is the center of mass energy and $\theta$ is the azimuthal angle of the scattering. 
The dependence from $\cos\theta$ can be eliminated by projecting the amplitude onto partial waves of total angular momentum
$J$. In our case it suffices to consider the lowest partial wave, defined by
\beq 
\label{PWprojprev}
a^0_{fi} = \frac{1}{32 \pi s} \int_{-1}^{1}
d(\cos\theta) \, \mathcal{M}_{fi} (\sqrt{s},\cos\theta) \, .
\eeq
This expression is only valid in the high-energy limit, since we neglected kinematical factors ensuring that the 
partial wave is zero at threshold (see e.g.~\cite{DiLuzio:2016sur}). 
The right hand side of \eq{PWprojprev} must be further multiplied by a $\frac{1}{\sqrt{2}}$ factor 
for any identical pair of particles either in the initial or final state. 
The unitarity of the $S$-matrix implies 
\beq
\label{unitS}
\frac{1}{2i} \left( a^0_{fi} - a^{0*}_{if} \right) \geq \sum_h a^{0*}_{hf} a^{0}_{hi} \, ,
\eeq
where the inequality originates from the fact that we restricted the sum over $h$ to 2-particle states. 
For $i=f$ \eq{unitS} reduces to 
$\Im a^0_{ii} \geq |a^0_{ii}|^2$ or, equivalently, 
$|\Im a^0_{ii} | \leq 1$ and $|\Re a^0_{ii} | \leq \frac{1}{2}$. 
It is customary to define the perturbative unitarity bound 
\beq
\label{UnitBound}
|\Re (a^0_{ii})^{\text{Born}} | \leq \frac{1}{2} \, , 
\eeq
at the level of the Born amplitude. Although the choice in \eq{UnitBound} 
is somewhat arbitrary, it yields a reasonable indication of the range of validity of the 
perturbative expansion. In fact, a Born value of $\Re a^0_{ii} = \frac{1}{2}$ and $\Im a^0_{ii} = 0$ 
needs at least a higher-order correction of $40\%$ in order to restore unitarity (see e.g.~\cite{DiLuzio:2016sur}), 
thus signalling the breakdown of the expansion itself. 

It is also useful to note that in order to optimize the unitarity bound one can look for correlations in the partial-wave matrix 
(e.g.~in the gauge group or flavour space). This corresponds to diagonalizing the partial-wave matrix and setting the bound on the 
largest eigenvalue, $\Re \tilde a^0_{ii} < 1/2$, with the forward scattering $i=f$ understood to correspond to 
a superposition of states which is an eigenvector of $\tilde a^0_{ii}$. 
Note that neglecting a scattering channel for the partial-wave matrix corresponds to removing the 
associated row/column. Thanks to the Cauchy interlacing theorem, we also know that the largest eigenvalue of the reduced 
matrix is always $\leq$ than the largest eigenvalue of the full matrix. Hence, by neglecting a scattering channel the unitarity 
bound still (conservatively) applies. 

\section{Unitarity bounds in the EFT}
\label{semilepSMEFT}

In this section we derive the connection between the scale of unitarity violation $\Lambda_U$ 
and the coefficients $\Lambda_{\mathcal{O}}$ of the semi-leptonic SMEFT basis, relevant for the $R_{D^{(*)}}$ and $R_{K^{(*)}}$ anomalies. 
At energies $\sqrt{s} \gg v$ the scattering amplitudes are conveniently described by 
exploiting the full SM invariance. 
A complete basis of semi-leptonic $d=6$ operators invariant under the SM gauge symmetry is \cite{Alonso:2015sja}
\begin{align}
\label{EFTbasissemil}
\mathcal{L}_{\rm SMEFT} 
& \supset \frac{1}{\Lambda^2_{{QL^{(3)}}}} (\bar Q_L \gamma_\mu \sigma^A Q_L) (\bar L_L \gamma^\mu \sigma^A L_L) 
+ \frac{1}{\Lambda^2_{{QL^{(1)}}}} (\bar Q_L \gamma_\mu Q_L) (\bar L_L \gamma^\mu L_L) 
+ \frac{1}{\Lambda^2_{{ue}}} (\bar u_R \gamma_\mu u_R) (\bar e_R \gamma^\mu e_R) \nonumber \\
& + \frac{1}{\Lambda^2_{{de}}} (\bar d_R \gamma_\mu d_R) (\bar e_R \gamma^\mu e_R) 
+ \frac{1}{\Lambda^2_{{uL}}} (\bar u_R \gamma_\mu u_R) (\bar L_L \gamma^\mu L_L) 
+ \frac{1}{\Lambda^2_{{dL}}} (\bar d_R \gamma_\mu d_R) (\bar L_L \gamma^\mu L_L)  \nonumber \\
& + \frac{1}{\Lambda^2_{{Qe}}} (\bar Q_L \gamma_\mu Q_L) (\bar e_R \gamma^\mu e_R) 
+ \frac{1}{\Lambda^2_{{dQLe}}} (\bar d_R \, Q_L) (\bar L_L \, e_R) + \frac{1}{\Lambda^2_{{QuLe}}} (\bar Q_L \, u_R) i \sigma^2 (\bar L_L \, e_R) \nonumber \\
& + \frac{1}{\Lambda^2_{{QuLe'}}} (\bar Q_L \sigma_{\mu\nu} u_R) i \sigma^2 (\bar L_L \sigma_{\mu\nu}  e_R) + \text{h.c.} \, ,
\end{align}
where flavour indices have been suppressed. Here, $Q_L$ and $L_L$ denote $SU(2)_L$ doublets, while 
$u_R$, $d_R$ and $e_R$ are $SU(2)_L$ singlets. 

An important aspect to be taken into account for the determination of the unitarity bound 
is the correlation of the scattering amplitude in the $SU(3)_C \times SU(2)_L$ space.  
Let us consider, for instance, the scattering $(Q_L)^\alpha_a  + (\bar Q_L)^\beta_b \to (L_L)_c +  (\bar L_L)_d$, 
where greek (latin) indices run over the fundamental of $SU(3)_C$ ($SU(2)_L$). 
Assuming a color singlet channel (which applies to all the operators in \eq{EFTbasissemil}) 
the amplitude in color space can be represented by a $4 \times 4$ matrix in the basis 
$\{ (Q_L)^1 (\bar Q_L)^1, (Q_L)^2 (\bar Q_L)^2, (Q_L)^3 (\bar Q_L)^3, (L_L) (\bar L_L) \}$.  
Similarly, in $SU(2)_L$ space we can represent it via a $4 \times 4$ matrix in the 
basis $\{ \psi_1 \bar \psi_1, \psi_1 \bar \psi_2, \psi_2 \bar \psi_1, \psi_2 \bar \psi_2 \}$, where 
$\psi_a$ ($a=1,2$ being an $SU(2)_L$ index) denotes 
either $(Q_L)^\alpha$ or $L_L$. 
A stronger unitarity bound can be hence obtained by preparing the initial and final states of the scattering 
in the eigenstate corresponding to the highest eigenvalue of $a^0$ both in 
$SU(3)_C$ and $SU(2)_L$ space (cf.~also the discussion at the end of \sect{UBs}). 
By looking at the different scattering channels displayed in the first column of \Table{UBEFTbasis}, 
we obtain for each case the scale of unitarity violation $\Lambda_U$ (defined as the value of $\sqrt{s}$ where 
the condition in \eq{UnitBound} is saturated) as a function of the 
scale of the SMEFT operator $\Lambda_{\mathcal{O}}$, where $\mathcal{O} = \{ QL^{(3)}, QL^{(1)}, \dots \}$. 
In the last column of \Table{UBEFTbasis} we also show the enhancement of the $a^0$ eigenvalue 
due to the $SU(3)_C \times SU(2)_L$ group structure of the partial wave. 
The full calculation of the unitarity bound for the triplet operator $\mathcal{O} = QL^{(3)}$ 
(including a detailed discussion of the gauge group enhancement) 
is exemplified in \app{tripop}, while the bounds for the other cases are obtained in a similar way. 
We finally observe that since the tensor operator does not contribute to the $J=0$ partial wave, in order to apply the 
unitarity bound from $\Lambda_{{QuLe'}}$ one would need to inspect higher partial waves. 
\begin{table}[htp]
\begin{center}
\begin{tabular}{|c|c|c|}
\hline
Scattering & $\Lambda_U$ & $SU(3)_C \times SU(2)_L$ \\
\hline
\hline
$(Q_L + \bar Q_L)_3 \to (L_L + \bar L_L)_3$ & $\sqrt{\frac{4 \pi}{\sqrt{3}}} \abs{\Lambda_{{QL^{(3)}}}}$ & $\sqrt{3} \times 2$ \\
$(Q_L + \bar Q_L)_1 \to (L_L + \bar L_L)_1$ & $\sqrt{\frac{4 \pi}{\sqrt{3}}} \abs{\Lambda_{{QL^{(1)}}}}$ & $\sqrt{3} \times 2$ \\
$u_R + \bar u_R \to e_R + \bar e_R$ & $\sqrt{\frac{8 \pi}{\sqrt{3}}} \abs{\Lambda_{{ue}}}$ & $\sqrt{3}$ \\
$d_R + \bar d_R \to e_R + \bar e_R$ & $\sqrt{\frac{8 \pi}{\sqrt{3}}} \abs{\Lambda_{{de}}}$ & $\sqrt{3}$ \\
$u_R + \bar u_R \to L_L + \bar L_L$ & $\sqrt{\frac{8 \pi}{\sqrt{6}}} \abs{\Lambda_{{uL}}}$ & $\sqrt{3} \times \sqrt{2}$ \\
$d_R + \bar d_R \to L_L + \bar L_L$ & $\sqrt{\frac{8 \pi}{\sqrt{6}}} \abs{\Lambda_{{dL}}}$ & $\sqrt{3} \times \sqrt{2}$ \\
$Q_L + \bar Q_L \to e_R + \bar e_R$ & $\sqrt{\frac{8 \pi}{\sqrt{6}}} \abs{\Lambda_{{Qe}}}$ & $\sqrt{3} \times \sqrt{2}$ \\
$d_R + \bar Q_L \to L_L + \bar e_R$ & $\sqrt{\frac{8 \pi}{\sqrt{3}}} \abs{\Lambda_{{dQLe}}}$ & $\sqrt{3}$ \\
$Q_L + \bar u_R \to L_L + \bar e_R$ & $\sqrt{\frac{8 \pi}{\sqrt{3}}} \abs{\Lambda_{{QuLe}}}$ & $\sqrt{3}$ \\
\hline
\end{tabular}
\end{center}
\caption{Scale of unitarity violation $\Lambda_U$ as a function of the coefficients $\Lambda_{\mathcal{O}}$ 
of the semi-leptonic SMEFT basis of \eq{EFTbasissemil}. For the case of $Q_L \bar Q_L \to L_L \bar L_L$ scattering 
the $SU(2)_L$ triplet and singlet channels are labelled explicitly. 
The third column denotes the enhancement factors  
on the partial wave due to the gauge group structure in $SU(3)_C \times SU(2)_L$ space.}
\label{UBEFTbasis}
\end{table}%


\section{Unitarity bounds in simplified models}
\label{simplmodels}

We continue by applying unitarity constraints on the parameter space of simplified models for the explanation of the 
$R_{D^{(*)}}$ and $R_{K^{(*)}}$ anomalies. Note that this case is slightly different from the unitarity bounds in the EFT, since 
the scattering amplitudes do not grow with the energy. Still, one can examine the $2 \to 2$ scatterings of 
SM fermions in order to set perturbativity limits on the renormalizable couplings of the new mediators and, in turn, translate them  
into an upper bound on the mass of the new states (once the ratio coupling/mass is fixed in terms of the fit to the relevant observable). 
As two representative classes of simplified models, we consider colorless spin-1 mediators and scalar/vector leptoquarks. 

However, some comments are in order about the phenomenological viability of the simplified models. 
The criterium that we are going to follow in order to select the suitable representations for the new mediators is that 
after integrating them out they are able to generate triplet and singlet left-handed operator, namely 
those associated with the coefficients $\Lambda_{QL^{(3)}}$ and $\Lambda_{QL^{(1)}}$ in \eq{EFTbasissemil}. 
In all the cases that we are going to consider the phenomenologically disfavoured right-handed and scalar/tensor operator 
of \eq{EFTbasissemil} can be set to zero by a proper choice of the mediator's coupling. 
Given these conditions, the full set of simplified models is displayed in \Table{simplmodel}.   
\begin{table}[htp]
\begin{center}
\begin{tabular}{|c|c|c|c|c|c|c|}
\hline
Simplified Model & Spin & SM irrep & $c_1 /c_3$ & $R_{D^{(*)}}$ & $R_{K^{(*)}}$ & No $d_i \to d_j \nu \bar \nu$ \\
\hline
\hline
$Z'$ & 1 & $(1,1,0)$ & $\infty$ & $\times$ & $\checkmark$ & $\times$ \\
$V'$ & 1 & $(1,3,0)$ & $0$ & $\checkmark$ & $\checkmark$ & $\times$ \\
$S_1$ & 0 & $(\bar 3,1,1/3)$ & $-1$ & $\checkmark$ & $\times$ & $\times$ \\
$S_3$ & 0 & $(\bar 3,3,1/3)$ & $3$ & $\checkmark$ & $\checkmark$ & $\times$ \\
\rowcolor{LightCyan} 
$U_1$ & 1 & $(3,1,2/3)$ & $1$ & $\checkmark$ & $\checkmark$ & $\checkmark$ \\
$U_3$ & 1 & $(3,3,2/3)$ & $-3$ & $\checkmark$ & $\checkmark$ & $\times$ \\
\hline
\end{tabular}
\end{center}
\caption{Overview of simplified models which can possibly contribute to $R_{D^{(*)}}$ or $R_{K^{(*)}}$ via a singlet/triplet 
left-handed operator. Only for specific values of the ratio of the Wilson coefficients 
$c_1 / c_3$ 
(obtained by integrating out a given mediator) 
the dangerous $d_i \to d_j \nu \bar \nu$ operators are not generated ($U_1$ case). 
}
\label{simplmodel}
\end{table}%

From the $SU(2)_L$ decomposition (neglecting flavour indices and reinserting the Wilson coefficients explicitly)
\begin{multline}
\label{tripsingdec}
\frac{c_1}{\Lambda^2} (\bar Q_L \gamma_\mu Q_L) (\bar L_L \gamma^\mu L_L) 
+ \frac{c_3}{\Lambda^2} (\bar Q_L \gamma_\mu \sigma^A Q_L) (\bar L_L \gamma^\mu \sigma^A L_L)
\\
= 
\frac{c_1 + c_3}{\Lambda^2} 
\left[ 
\left(\bar d_L \gamma_\mu d_L) (\bar e_L \gamma^\mu e_L\right) + 
\left(\bar u_L \gamma_\mu u_L) (\bar \nu_L \gamma^\mu \nu_L\right)
\right] \\
+ \frac{c_1 - c_3}{\Lambda^2} 
\left[ \left(\bar d_L \gamma_\mu d_L) (\bar \nu_L \gamma^\mu \nu_L\right) 
+  \left(\bar u_L \gamma_\mu u_L) (\bar e_L \gamma^\mu e_L\right) \right] \\
+ 2 \frac{c_3}{\Lambda^2}
\left[ \left(\bar u_L \gamma_\mu d_L) (\bar e_L \gamma^\mu \nu_L\right) 
+  \left(\bar d_L \gamma_\mu u_L) (\bar \nu_L \gamma^\mu e_L\right) \right] 
\, ,
\end{multline}
it is evident that for $c_1 / c_3 = -1$ there are 
no $b \to s \mu \bar \mu$ transitions. Similarly, for $c_1 / c_3 = 1$ 
processes of the type $d_i \to d_j \nu \bar \nu$ are absent. The latter are particularly dangerous, 
since decays like $B \to K^{(*)} \nu \bar \nu$ or $K \to \pi \nu \bar \nu$ are very constraining \cite{Buras:2014fpa,Bordone:2017lsy}. 
From this point of view $U_1$ is phenomenologically favoured, since it automatically ensures the absence 
of $d_i \to d_j \nu \bar \nu$ operators at the scale of the 
threshold.\footnote{This can also be achieved in non-minimal scenarios with two leptoquarks via 
a proper cancellation \cite{Crivellin:2017zlb}.} 
For an incomplete list of references addressing both $R_{D^{(*)}}$ and $R_{K^{(*)}}$ 
with this leptoquark see \cite{Calibbi:2015kma,Barbieri:2015yvd,Barbieri:2016las,DiLuzio:2017vat}.  
Other phenomenological issues that have to be taken into account when considering a simplified model 
are electroweak precision tests and the radiative generation of 
LFU breaking effects in $Z$ and $\tau$ decays \cite{Feruglio:2016gvd,Feruglio:2017rjo}.    
In order to avoid those bounds one has to assume either a certain level of tuning within the couplings of the 
simplified model or rely on some non-generic features of the UV completion of the simplified model. 
For an example of a leptoquark model where all these bounds have been consistently addressed see e.g.~\cite{Altmannshofer:2017poe}. 
Finally, one has to consider direct searches that we briefly address in \sect{lhc}. Our results on the 
unitarity bounds for colorless vectors and leptoquarks, which are summarized in \Tables{tablecolorlessspin1}{tableLQ}, 
provide an extra constraint which has to be satisfied within perturbative models.

\subsection{Colorless vectors}
\label{colorlessvec}

Let us first consider the case of a real electroweak vector, $V'_\mu \sim (1,3,0)$, which couples to the SM fermions via 
\beq 
\label{LWp}
\mathcal{L}_{V'} \supset 
\lambda^Q_{ij} \, \bar Q_i \gamma^\mu \sigma^A Q_j V'^A_\mu + 
\lambda^L_{ij} \, \bar L_i \gamma^\mu \sigma^A L_j V'^A_\mu + \text{h.c.} \, .
\eeq
At energies $\sqrt{s} \gg M_{V'}$ the partial-wave scattering matrix in the $(Q_j \bar Q_i, L_l \bar L_k)$ basis is given by\footnote{An extra 
channel with $V' V'$ in the initial/final state opens up at energies $\sqrt{s} > 2 M_{V'}$. By neglecting such contribution, 
the unitarity bound obtained by considering the reduced partial-wave matrix conservatively applies (cf.~the discussion at the end of 
\sect{UBs}).} 
\beq 
\label{a0WpEFT}
a^0 = \frac{1}{8\pi} 
\left( 
\begin{array}{cc}
3 |\lambda^Q_{ij}|^2 & \sqrt{3} \lambda^Q_{ij} (\lambda^L_{kl})^\ast \\
\sqrt{3} (\lambda^Q_{ij})^\ast \lambda^L_{kl} & |\lambda^L_{kl}|^2
\end{array}
\right) \, , 
\eeq
where we also took into account the $SU(3)_C \times SU(2)_L$ multiplicity factors. The formalism for extracting the 
correlation in the gauge group space follows very closely the sample calculation of the scattering with the effective triplet operator, 
which is detailed in \app{tripop}. 
The largest eigenvalue of \eq{a0WpEFT} is
\beq 
a^0 = \frac{3 |\lambda^Q_{ij}|^2 + |\lambda^L_{kl}|^2}{8\pi} \, , 
\eeq
and the associated unitarity bound reads 
\beq 
\label{pertboundgQgL}
3 |\lambda^Q_{ij}|^2 + |\lambda^L_{kl}|^2 < 4 \pi \, .
\eeq
Note that this is stronger than the perturbativity bound sometimes quoted in the literature, 
e.g.~$|\lambda^{Q,L}_{ij}| < \sqrt{4 \pi}$ \cite{Greljo:2015mma}. 
In the following, we exemplify the unitarity bounds in the case where the couplings of $V'$ are aligned 
with the operators responsible for $R_{D^{(*)}}$ and $R_{K^{(*)}}$, respectively 
$\lambda^Q_{23} \lambda^L_{33}$ and $\lambda^Q_{23} \lambda^L_{22}$. 
This actually yields the most conservative bounds 
without flavour enhancements. The generalization to non-aligned cases is straightforward and it is reported 
in \Table{tablecolorlessspin1} for some representative cases. 
In this respect, we note that the multiple coupling configuration with 
$\lambda^Q_{33} \sim \lambda^Q_{23}$ might help in relaxing the bounds from Refs.~\cite{Feruglio:2016gvd,Feruglio:2017rjo}.
Integrating out the $V'$ and matching with \eq{LeffRDRK}, we obtain
\beq 
\label{matchWp}
\frac{\lambda^Q_{23} \lambda^L_{33}}{M^2_{V'}} = \frac{1}{\Lambda^2_{R_{D^{(*)}}}} \, , 
\qquad 
-\frac{\lambda^Q_{23} \lambda^L_{22}}{M^2_{V'}} = \frac{1}{\Lambda^2_{R_{K^{(*)}}}} \, .
\eeq
It is convenient to define the auxiliary functions 
\beq 
r = \abs{\frac{\lambda^L_{kl}}{\lambda^Q_{ij}}} \qquad \text{and} \qquad f(r) = \frac{r}{3 + r^2} \, ,
\eeq
so that the bound in \eq{pertboundgQgL} can be recast as (using also \eq{matchWp})
\beq 
M_{V'} < \sqrt{4 \pi f(r)} \Lambda_A \, ,  
\eeq
where $A = \{R_{D^{(*)}}, R_{K^{(*)}}\}$.
The most conservative bound is obtained by maximizing the function $f(r)$ at $r = \sqrt{3}$, which yields 
\beq 
M_{V'} < \sqrt{\frac{2 \pi}{\sqrt{3}}} \Lambda_A = 6.5 \ \text{TeV} \ (59 \ \text{TeV}) \, ,
\eeq
for the case of $R_{D^{(*)}}$ ($R_{K^{(*)}}$). 

The analysis for the $Z'$ is basically identical to that of the $V'$ as far as concerns neutral currents. 
So we do not repeat it here. The unitarity bounds for both the cases are collected in \Table{tablecolorlessspin1}.    

\begin{table}[htp]
\begin{center}
\begin{tabular}{|c|c|c|c|c|c|}
\hline
Anomaly & Coupling & FS$_{Q}$  & FS$_{L}$ & $M_{V'} [\text{TeV}]$ & $M_{Z'} [\text{TeV}]$ \\
\hline
\hline
$ b \to c \tau \bar \nu $ & $\lambda^Q_{23}\,\lambda^L_{33}$ & 1 & 1 & 6.5 & $\times$ \\
$ b \to c \tau \bar \nu $ & $\lambda^Q_{33}\,\lambda^L_{33}$ & $\abs{V_{cb}}$ & 1 & 1.3 & $\times$ \\
\hline
$ b \to s \mu \bar \mu $ & $\lambda^Q_{23}\,\lambda^L_{22}$ & 1 & 1 & 59 & 59 \\
$ b \to s \mu \bar \mu $ & $\lambda^Q_{33}\,\lambda^L_{22}$ & $\abs{V_{ts}}$ & 1 & 12 & 12  \\
$ b \to s \mu \bar \mu $ & $\lambda^Q_{33}\,\lambda^L_{33}$ & $\abs{V_{ts}}$ & $m_{\mu}/m_{\tau}$& 2.9 & 2.9  \\
$ b \to s \mu \bar \mu $ & $\lambda^Q_{33}\,\lambda^L_{33}$ & $\abs{V_{ts}}$ & $(m_{\mu}/m_{\tau})^2$& 0.7 & 0.7 \\
\hline
\end{tabular}
\end{center}
\caption{Summary of unitarity bounds for colorless spin-1 mediators. 
FS$_{Q}$ and FS$_{L}$ denote the flavour suppression factors in the quark and lepton sectors 
(same as in \Table{tablescales}).
}
\label{tablecolorlessspin1}
\end{table}%

\subsection{Leptoquarks}
\label{leptoq}

Let us start by first discussing the flavour structure of the leptoquark Lagrangian. 
Neglecting Lorentz and gauge indices, we have 
$\mathcal{L}_{LQ} \supset y^{ij}_{QL} Q_i L_j \Phi + \text{h.c.}$,
where $\Phi$ denotes one of the four leptoquarks in \Table{simplmodel}. 
The simplest way to generate a contribution for either $R_{D^{(*)}}$ or $R_{K^{(*)}}$ is 
to switch on a single coupling, e.g.~$y_{QL}^{3 j}$, with the lepton index $j$ aligned either along the third 
or second generation. The $3 \to 2$ transition in the quark sector can be then obtained either via a
$V_{cb}$ or $V_{ts}$ suppression. However, within such an approach the sign of the Wilson coefficient, 
which goes either like $\abs{y^{3j}_{QL}}^2 V_{cb}$ or $\abs{y^{3j}_{QL}}^2 V_{ts}$ (recall that 
$V_{cb}>0$ and $V_{ts}<0$ in the standard parametrization), is fixed and does not always correspond 
to the one necessary to reproduce the anomaly.\footnote{With the 
single coupling $y_{QL}^{3 j}$ we find that $S_3$ cannot explain neither of the anomalies, while 
$U_3$ cannot explain $R_{D^{(*)}}$.} Hence, in the following we define our simplified models 
based on the two leptoquark couplings $y_{QL}^{3 j}$ and $y_{QL}^{2 j}$, 
so that the sign of the contribution can be always matched.  
We further assume the scaling $y_{QL}^{2 j} \sim y_{QL}^{3 j} \lambda^2$, 
as suggested by motivated flavour structures. 
Given the hierarchy $y_{QL}^{3 j} \gg y_{QL}^{2 j}$, the scattering amplitudes 
are dominated by $y_{QL}^{3 j}$ and the correlation of the partial wave in flavour 
space can be safely neglected. Indeed, for a leptoquark-mediated processes in the $t$-channel 
one should make the following replacement in the bound: $|y_{QL}^{3j}|^2 \to 
\sqrt{|y_{QL}^{3j}|^4 + |y_{QL}^{2j}|^4} \sim |y_{QL}^{3j}|^2  \sqrt{1 + \lambda^4}$, 
while no such flavour enhancement is even present for an $s$-channel scattering. 
Given these considerations, for each leptoquark of \Table{UBEFTbasis} 
we compute the unitarity constraints on its couplings and the matching 
condition with the effective operators in \eq{LeffRDRK}. Following the conventions of Ref.~\cite{Dorsner:2016wpm} we have:
\begin{itemize}
\item $S_1 \sim (\bar 3,1,1/3)$: \
$
\mathcal{L}_{S_1} \supset 
y^{ij}_{QL} \overline{Q^c}_{i,a} \epsilon^{ab} L_{j,b} \, S_1 + \text{h.c.}
$

The strongest unitarity bound comes from the $t$-channel mediated $Q^c_3 \overline{Q^c}_3 \to L_j \bar L_j$ scattering, 
which in the limit $\sqrt{s} \gg M_{S_1}$ gives 
\beq 
\abs{y^{3j}_{QL}}^2 < \frac{8 \pi}{\sqrt{3}} \, , 
\eeq
where we included a $\sqrt{3}$ enhancement factor due to the correlation 
of the partial wave in color space.
Integrating out $S_1$ and matching with the operators in \eq{LeffRDRK} we obtain 
\beq 
\frac{\abs{y^{33}_{QL}}^2 \lambda^2}{2 M^2_{S_1}} = \frac{1}{\Lambda^2_{R_{D^{(*)}}}} \, .
\eeq

\item $S_3 \sim (\bar 3,3,1/3)$: \
$\mathcal{L}_{S_3} \supset 
y^{ij}_{QL} \overline{Q^c}_{i,a} (\epsilon\sigma^A)^{ab} L_{j,b} \, S^A_3  + \text{h.c.}$

Analogously to the previous case we consider the $t$-channel mediated $Q^c_3 \overline{Q^c}_3 \to L_j \bar L_j$ scattering,  
from which we get the unitarity bound
\beq 
\label{UBS1}
\abs{y^{3j}_{QL}}^2 < \frac{8 \pi}{3 \sqrt{3}} \, ,  
\eeq
where we included a $\sqrt{3} \times 3$ enhancement factor due to the correlation 
of the partial wave in the $SU(3)_C \times SU(2)_L$ space.
Integrating out $S_3$ and matching with the operators in \eq{LeffRDRK} we obtain 
\beq 
\frac{\abs{y^{33}_{QL}}^2 \lambda^2}{2 M^2_{S_3}} = \frac{1}{\Lambda^2_{R_{D^{(*)}}}} 
\qquad \text{and} \qquad 
\frac{\abs{y^{32}_{QL}}^2 \lambda^2}{M^2_{S_3}} = \frac{1}{\Lambda^2_{R_{K^{(*)}}}} \, .
\eeq

\item $U_1 \sim (3,1,2/3)$: \ 
$\mathcal{L}_{U_1} \supset 
y^{ij}_{QL} \bar Q_{i,a} \gamma^\mu \delta^{ab} L_{j,b}  U_{1,\mu} + \text{h.c.}$

By examining the $s$-channel process $\bar Q_3 L_j \to \bar Q_3 L_j$ at $\sqrt{s} \gg M_{U_1}$ we extract the unitarity 
bound\footnote{The $t$-channel mediated $Q_3 \bar Q_3 \to L_j \bar L_j$ scattering 
cannot be straightforwardly used here, since the $J=0$ partial wave is formally divergent. 
This is due to the Coulomb singularity in the forward direction of the scattering for $\sqrt{s} \gg M_{U_1}$.} 
\beq 
\label{UBx1}
\abs{y^{3j}_{QL}}^2 < 4 \pi \, ,  
\eeq
where we included a factor 2 enhancement from the correlation of the partial wave in the $SU(2)_L$ space 
(while there is no $SU(3)_C$ enhancement since the color flows through the diagram). 
Integrating out $U_1$ and matching with the operators in \eq{LeffRDRK} we obtain 
\beq 
\frac{\abs{y^{33}_{QL}}^2 \lambda^2}{M^2_{U_1}} = \frac{1}{\Lambda^2_{R_{D^{(*)}}}} 
\qquad \text{and} \qquad 
\frac{\abs{y^{32}_{QL}}^2 \lambda^2}{M^2_{U_1}} = \frac{1}{\Lambda^2_{R_{K^{(*)}}}} \, .
\eeq

\item $U_3 \sim (3,3,2/3)$: \ 
$\mathcal{L}_{U_3} \supset 
y^{ij}_{QL} \bar Q_{i,a} \gamma^\mu (\sigma^A)^{ab} L_{j,b}  U^A_{3,\mu} + \text{h.c.}$ 

Analogously to the previous case, from the $s$-channel process $\bar Q_3 L_j \to \bar Q_3 L_j$ we obtain 
\beq 
\abs{y^{3j}_{QL}}^2 < 4 \pi \, .
\eeq
Integrating out $U_3$ and matching with the operators in \eq{LeffRDRK} we obtain 
\beq 
\frac{\abs{y^{33}_{QL}}^2 \lambda^2}{M^2_{U_3}} = \frac{1}{\Lambda^2_{R_{D^{(*)}}}} 
\qquad \text{and} \qquad 
\frac{\abs{y^{32}_{QL}}^2 \lambda^2}{M^2_{U_3}} = \frac{1}{\Lambda^2_{R_{K^{(*)}}}} \, .
\eeq
\end{itemize}
After saturating the matching condition required to reproduce 
the anomalies, we can translate the unitarity bounds on the leptoquark couplings 
into an upper bound on the leptoquark masses. 
As a reference value we fix $\lambda^2 = \abs{V_{cb}}$ ($\abs{V_{ts}}$) for $R_{D^{(*)}}$ ($R_{K^{(*)}}$). 
The results are displayed in \Table{tableLQ}, depending on the flavour structure of the leptoquark couplings.  
\begin{table}[htp]
\begin{center}
\begin{tabular}{|c|c|c|c|c|c|c|c|}
\hline
Anomaly & Coupling & FS$_{Q}$  & FS$_{L}$ & $M_{S_1} [\text{TeV}]$ & $M_{S_3} [\text{TeV}]$ & $M_{U_1} [\text{TeV}]$ & $M_{U_3} [\text{TeV}]$ \\
\hline
\hline
$ b \to c \tau \bar \nu $ & $y_{QL}^{33}$ & $\abs{V_{cb}}$ & 1 & $1.3$ & \cellcolor{mediumcandyapplered} $0.8$ & 1.7 & 1.7 \\
\hline
$ b \to s \mu \bar \mu $ & $y_{QL}^{32}$ & $\abs{V_{ts}}$ & 1 & $\times$ & 14 & 22 & 22 \\
$ b \to s \mu \bar \mu $ & $y_{QL}^{33}$ & $\abs{V_{ts}}$ & $m_{\mu}/m_{\tau}$& $\times$ & 3.3 & 5.4 & 5.4 \\
$ b \to s \mu \bar \mu $ & $y_{QL}^{33}$ & $\abs{V_{ts}}$ & $(m_{\mu}/m_{\tau})^2$ & $\times$ & \cellcolor{mediumcandyapplered} 0.8 & \cellcolor{piggypink} 1.3 & \cellcolor{piggypink} 1.3 \\
\hline
\end{tabular}
\end{center}
\caption{Summary of unitarity bounds for leptoquarks. 
FS$_{Q}$ and FS$_{L}$ indicate the flavour suppression factors in the quark and lepton sectors (same as in \Table{tablescales}). 
Red (light-red) boxes denote the cases excluded (disfavoured) by direct searches (see \sect{lhc}).  
}
\label{tableLQ}
\end{table}%

\subsection{Direct searches at the LHC}
\label{lhc}

We will now briefly discuss the bounds from direct searches for the simplified models 
of \Table{simplmodel} and compare them with the unitarity bounds on the new mediators' 
masses from \Tables{tablecolorlessspin1}{tableLQ}. 
We will focus in particular on decay channels involving the third family, 
since these are theoretically motivated by flavour models and because it is precisely in those cases that the upper bounds 
on the mass of the new states are more stringent. 

Let us discuss in turn the various cases. Ref.~\cite{Faroughy:2016osc} considered vector triplet $V'$ 
exclusion limits by recasting $pp (b\bar b) \to \tau \bar \tau$ searches. The conclusion is that for 
relatively heavy vectors $M_{V'} \gtrsim 500$ GeV the resolution of the $R_{D^{(*)}}$ anomaly 
with dominant third generation couplings requires a very large $Z'$ decay width 
(where here $Z'$ denotes here the neutral component of $V'$), which is beyond the perturbative regime. 
This is somehow compatible with our unitarity bound $M_{V'} < 1.3$ TeV in \Table{tablecolorlessspin1}. 
Note, however, that a large $Z'$ width also implies extra model-dependent decay channels which 
would in principle contribute to our scattering amplitudes, and would yield in turn a bound stronger than $1.3$ TeV. 
On the other hand, for light masses $M_{Z'} \lesssim 400$ GeV a perturbative window 
with a relatively small $Z'$ width is not yet excluded by $\tau \bar \tau$ searches. 
However, this requires a suppression of electroweak precision observables which are generically 
quite constraining \cite{Pappadopulo:2014qza}. 

If leptoquarks are light enough they can be pair-produced at LHC with sizable cross-section via QCD interactions. 
As already stated we assume that decay channels are dominated by third generation SM fermions. 
$S_1$ has the same quantum numbers of a sbottom and decays into either 
$S_1 \to \bar b \bar \nu_\tau$ or $S_1 \to \bar t \bar \tau$ (both with $\mathcal{B}=50 \%$). 
Using the results of \cite{Gripaios:2014tna} we obtain $M_{S_1} > 570$ GeV, 
which is still compatible with the unitarity bound in \Table{tableLQ}. 
On the other hand, $S_3$ comprises three charge eigenstates, 
respectively with charges $4/3$, $1/3$ and $-2/3$. The predominant decays are 
$S_3^{4/3} \to \bar b \bar \tau$,
$S_3^{1/3} \to \bar b \bar \nu_\tau$ or $S_3^{1/3} \to \bar t \bar \tau$ (both with $\mathcal{B}=50 \%$) 
and $S_3^{-2/3} \to \bar t \bar \nu_\tau$. There will be electroweak mass splittings between the three leptoquark states, allowing the heavier ones to decay to the lighter ones, but these decays will be subdominant to those through the leptoquark couplings, if the mass splittings are small. In fact, by using the results of Ref.~\cite{DiLuzio:2015oha} we find that electroweak precision data 
exclude mass splittings within $S_3$ above $\mathcal{O}(25)$ GeV. 
For $S_3^{-2/3}$ we can infer a bound of $M_{S_3^{-2/3}} \gtrsim 950$ GeV, by looking at 
SUSY searches for $\tilde t \to t \tilde \chi_0$ \cite{ATLAS:2017kyf}. For $S_3^{4/3}$ there is 
a dedicated leptoquark search for third generation final states \cite{Sirunyan:2017yrk}, 
which yields $S_3^{4/3} \gtrsim 850$ GeV. The bound on $S_3^{1/3}$ basically corresponds to the 
previous one for $S_1$. All in all, when comparing the limits from direct searches with the unitarity bounds in 
\Table{tableLQ}, we conclude that a leptoquark $S_3$ with couplings dominantly aligned along the third generation 
cannot explain within a perturbative framework the $R_{D^{(*)}}$ anomaly 
(and $R_{K^{(*)}}$ as well, under the hypothesis of MFV in the lepton sector).  

We finally discuss vector leptoquarks. Under the assumption of leading third generation 
couplings one can look at $pp (b\bar b) \to \tau \bar \tau$ searches, which however 
are not yet sensitive enough to rule out the explanation of $R_{D^{(*)}}$ 
via $U_1$ \cite{Faroughy:2016osc}. On the other hand, vector leptoquarks can also be efficiently pair-produced at LHC via 
their coupling to gluons. This interaction depends however on the UV completion of the vector. 
The most general CP-conserving Lagrangian describing the interaction of the 
vector $U_\mu$ with gluons (including operators up to $d=4$)
is given by \cite{Blumlein:1996qp}
\beq 
\mathcal{L}_{U}^g = - \frac{1}{2} \left( D_{[\mu} U_{\nu]} \right)^\dag D^{[\mu} U^{\nu]} 
+ M_{U}^2 U_\mu^\dag U^{\mu}
- i g_s (1 - \kappa_G) U^\dag_\mu t^a U_\nu G^{a\mu\nu} \, , 
\eeq
where $D_\mu = \partial_\mu - i g_s t^a G^a_\mu$ is the QCD covariant derivative 
and $G^a_{\mu\nu} = \partial_\mu G^a_\nu - \partial_\nu G^a_\mu + g_s f^{abc} G^b_\mu G^c_\nu$ 
is the usual QCD field strength. 
As two benchmark scenarios we consider the minimal coupling (MC) 
and the Yang Mills (YM) type of coupling of Ref.~\cite{Blumlein:1996qp}. The former case ($\kappa_G = 1$),
refers to the interaction stemming purely from the QCD covariant derivative of the vector, 
while the latter ($\kappa_G = 0$) includes non-minimal interactions 
between the vector and the gluons which arise when the vector has a gauge 
origin.\footnote{The explanation of $R_{D^{(*)}}$ and/or $R_{K^{(*)}}$ via gauge leptoquarks is strongly disfavoured. 
In fact, gauge invariance enforces extra constraints on the vector Lagrangian, like e.g.~the unitarity of the leptoquark interactions in flavour space (see \cite{Biggio:2016wyy} for a recent discussion).} We remark, however, 
that $\kappa_G$ is an unknown parameter. 

In the exact $U(2)$ flavour limit $U_1$ decays in either $U_1 \to t \bar \nu_\tau$ or 
$U_1 \to b \bar \tau$ (both with $\mathcal{B}=50 \%)$. 
By revisiting a $\sqrt{s} = 8$ TeV ATLAS search \cite{Aad:2015caa} for QCD pair-produced third generation scalar leptoquark in the $t \bar t \nu \bar \nu$ channel, Ref.~\cite{Barbieri:2015yvd} 
excludes $M_{U_1} < 770$ GeV. The latter exclusion actually applies to the MC scenario. 
In the meanwhile, there has been a new analysis at $\sqrt{s} = 13$ TeV \cite{Sirunyan:2017yrk}
for searches of scalar leptoquarks decaying in third generation SM fermions. 
We perform a rescaling of the bounds by employing the results in Ref.~\cite{Blumlein:1996qp} 
on the vector leptoquark total cross-section and extract the bounds: 
$M_{U_1} \gtrsim 1.0$ TeV (MC case) and $M_{U_1} \gtrsim 1.3$ TeV (YM case). 
Given the vicinity to the unitarity bounds in \Table{tableLQ},  
we remark that a dedicated experimental search in this case would be very helpful. 
Finally, since $U_3$ contains an isospin component with the same charge of $U_1$ 
we expect similar bounds, though optimized searches for the other charge eigenstates might yield 
better constraints.

\section{Conclusions}
\label{concl}

In this work we have investigated the constraints of partial-wave unitarity for the new physics interpretation 
of the recent hints of LFU violation in $B$-meson semi-leptonic decays, both within an EFT approach and by employing 
simplified models. In order to simplify the discussion we focussed on a single 
$SU(2)_L$ triplet operator (cf.~\eq{opQijL33}) which can contribute to both $R_{D^{(*)}}$ or $R_{K^{(*)}}$, 
but without necessarily relying on the common explanation of both the anomalies. 
This can be straightforwardly extended to the more general situation involving multiple operators, 
by employing the results of \sect{semilepSMEFT} in which we derived the connection between the scale of unitarity violation 
$\Lambda_U$ and the coefficients $\Lambda_{\mathcal{O}}$ of the semi-leptonic SMEFT operator basis. 

The results of the EFT analysis are summarized in \Table{tablescales}. 
In particular, we find that the most conservative bound on the scale of unitarity violation 
is $\Lambda_U = 9.2$ TeV and $84$ TeV, respectively for $R_{D^{(*)}}$ or $R_{K^{(*)}}$. This corresponds 
to the case when the effective operators are aligned in the direction of the flavour eigenstates responsible for the anomalies. 
On the other hand, motivated frameworks like e.g.~MFV, $U(2)$ flavour models and PC suggest 
an alignment of the effective operators along the third generation, thus implying that stronger unitarity bounds 
can be actually extracted by considering third generation fermions' scatterings. 
For instance, in the case of third generation alignment in the quark sector 
the previous bounds become $\Lambda_U = 1.9$ TeV and $17$ TeV, 
with the latter reaching even few TeV in the case of hierarchical flavour structures also in the lepton sector. 

In a similar way one can use the tool of perturbative unitarity to set constraints on the parameter space 
of simplified models explaining the $B$-flavour anomalies. 
As a representative class of models we considered colorless vectors and scalar/vector leptoquarks (cf.~\Table{simplmodel}).  
In all those cases, it was possible to use partial-wave unitarity in order to set bounds on the renormalizable 
couplings of the new mediators with the SM fermions. By fixing the ratio coupling/mass in order to fit the anomaly, 
the unitarity bound was hence translated into an upper bound on the mass of the simplified model's mediator. 
The results are collected in \Tables{tablecolorlessspin1}{tableLQ} for some reference flavour structures. 

While for the anomalies in $b \to s \ell \bar \ell$ transitions it is much easier to accomodate direct searches, 
that is not the case for $R_{D^{(*)}}$. Simplified models for explaining the latter are problematic for various reasons: 
$d_i \to d_j \nu \bar \nu$ transitions, electroweak precision observables, radiative generation of 
LFU breaking effects in $Z$ and $\tau$ decays, etc. 
On top of that, one should take into account unitarity constraints within perturbatively calculable models. 
The vector leptoquark $U_1$ seems phenomenologically in a better shape for explaining $R_{D^{(*)}}$, 
since it is automatically free from issues like $d_i \to d_j \nu \bar \nu$ transitions 
and also because, being an $SU(2)_L$ singlet, bounds from electroweak precision data are more easily evaded. 
For this specific case we provided a new bound by rescaling recent searches at 
LHC Run-2 with full dataset, finding that $M_{U_1} \gtrsim 1 \div 1.3$ TeV (depending on the UV completion of the vector).
In those cases where the leptoquark couplings are dominantly aligned along the third generation, 
the open window between direct searches and perturbativity is quite reduced 
and might be eventually closed in the near future.

\section*{Acknowledgments}

We thank Dario Buttazzo, Roberto Franceschini, Ramona Gr\"{o}ber, Jorge Martin Camalich, David Marzocca, Enrico Nardi, Giovanni Ridolfi,  
Alessandro Strumia and Riccardo Torre for helpful discussions, Jernej F.~Kamenik for collaboration in the early stages of this project 
and Gian Francesco Giudice for encouraging support. 

\appendix 

\section{Sample calculation: $SU(2)_L$ triplet operator}
\label{tripop}

In this Appendix we exemplify the calculation of the unitarity bound in the presence of the triplet operator 
\beq 
\label{Q3Q3L3L3op}
\frac{1}{\Lambda^2_{QL^{(3)}}} \left( \bar Q_L \gamma^\mu \sigma^A Q_L \right) \left( \bar L_L \gamma_\mu \sigma^A L_L \right) \, .  
\eeq
We are interested in evaluating the scattering amplitude 
\beq 
Q (p,r,a,\alpha) + \bar Q (k,s,b,\beta) \to L (p',r',c) + \bar L (k',s',d) \, ,
\eeq
where the indices $(p,r,a,\alpha)$ denote respectively momentum, polarization, $SU(2)_L$ and color indices. 
 The Lorentz invariant matrix element is given by
\beq 
\mathcal{M} = - \frac{1}{4\Lambda_{QL^{(3)}}^2} \delta_{\alpha\beta} (\sigma^A)_{ab}  (\sigma^A)_{cd} 
\left( \bar v^s(k) \gamma_\mu (1-\gamma_5) u^r(p) \bar u^{r'}(p') \gamma^\mu (1-\gamma_5) v^{s'}(k') \right) \, .
\eeq	
Since in the massless limit the fermions in \eq{Q3Q3L3L3op} are helicity 
eigenstates, at energies $\sqrt{s} \gg v$ only the $+--+$ polarization survives, yielding\footnote{We refer to Appendix A.2 of 
Ref.~\cite{DiLuzio:2016sur} for the explicit representation of the spinorial variables.} 
\beq 
\label{Mtripletop}
\mathcal{M}_{+--+} (\sqrt{s},\cos\theta) \overset{\sqrt{s} \gg v}{\simeq} \frac{2}{\Lambda_{QL^{(3)}}^2} \delta_{\alpha\beta} (\sigma^A)_{ab}  (\sigma^A)_{cd} \, 
s \cos^2\frac{\theta}{2} \, .
\eeq
The $J=0$ partial-wave scattering matrix is obtained via 
\beq 
\label{a0tripletop}
a^0 \overset{\sqrt{s} \gg v}{\simeq} \frac{1}{32} \int_{-1}^{+1} d(\cos\theta) \mathcal{M}_{+--+} (\sqrt{s},\cos\theta) 
= \frac{s}{16 \pi} \frac{1}{\Lambda_{QL^{(3)}}^2} \delta_{\alpha\beta} (\sigma^A)_{ab}  (\sigma^A)_{cd} \, .
\eeq
In order to maximize the unitarity bound one can prepare the scattering eigenstates in such a way that they 
correspond to the highest eigenvalues of $a^0$ in the gauge group space. 
Let us discuss in turn the $SU(3)_C$ and $SU(2)_L$ structures. 
In the former case the partial wave can be represented via the matrix 
(defined on the basis $\{ (Q_L)^1 (\bar Q_L)^1, (Q_L)^2 (\bar Q_L)^2, (Q_L)^3 (\bar Q_L)^3, (L_L) (\bar L_L) \}$)
\beq
a^0_{SU(3)_C} = 
\left(
\begin{array}{cccc}
0 & 0 & 0 & 1 \\
0 & 0 & 0 & 1 \\
0 & 0 & 0 & 1 \\
1 & 1 & 1 & 0 \\
\end{array}
\right) \, ,
\eeq
whose eigenvalues are $(\sqrt{3}, -\sqrt{3}, 0, 0)$. Thus, by preparing 
the initial and final states of the scattering in the eigenstate $\tfrac{1}{\sqrt{6}} (1,1,1,\sqrt{3})$, 
the color enhancement factor corresponds to $\sqrt{3}$. On the other hand, 
the partial wave in the $SU(2)_L$ space has the matrix form 
\beq
\label{a0SU2}
a^0_{SU(2)_L} = 
\left(
\begin{array}{cccc}
1 & 0 & 0 & -1 \\
0 & 0 & 2 & 0 \\
0 & 2 & 0 & 0 \\
-1 & 0 & 0 & 1 \\
\end{array}
\right) \, ,
\eeq
defined on the basis $\{ \psi_1 \bar \psi_1, \psi_1 \bar \psi_2, \psi_2 \bar \psi_1, \psi_2 \bar \psi_2 \}$, 
where $\psi_a$ ($a=1,2$ being an $SU(2)_L$ index) denotes either $(Q_L)^\alpha$ or $L_L$. 
In order to derive \eq{a0SU2} it is convenient to use the Fierz identity 
$(\sigma^A)_{ab}  (\sigma^A)_{cd} = 2 \delta_{ad} \delta_{cb} - \delta_{ab} \delta_{cd}$.
Since the eigenvalues of $a^0_{SU(2)_L}$ are $(2,2,-2,0)$, by preparing 
the initial and final states of the scattering in the eigenstate $\tfrac{1}{\sqrt{2}}(0,1,1,0)$, 
the $SU(2)_L$ enhancement factor is 2.
Summarizing, the gauge group enhancement leads to an extra $\sqrt{3} \times 2$ factor in the partial-wave eigenvalue, and
including the latter we obtain 
\beq 
a^0 = \frac{\sqrt{3}}{8 \pi} \frac{s}{\Lambda_{QL^{(3)}}^2} \, .  
\eeq
From the condition in \eq{UnitBound} it finally follows the unitarity bound $\sqrt{s} < \Lambda_U$, 
where
\beq 
\Lambda_U = \sqrt{\frac{4 \pi}{\sqrt{3}}} \abs{\Lambda_{QL^{(3)}}} \, . 
\eeq 

As a final remark, we briefly mention an alternative way to work out the gauge group enhancement
which employs irreducible representations for the scattering amplitude \cite{DiLuzio:2016sur}.  
Denoting by $\psi_i$ ($\overline{\psi}_j$) the fundamental (anti-fundamental) representation of an $SU(N)$ group, 
a general two-particle state $|\psi_i \overline{\psi}_j \rangle$ 
can be decomposed into a singlet and an adjoint channel
\begin{align}
& |\psi \overline{\psi}\rangle_\mathbf{1} 
= \frac{\delta_{ij}}{\sqrt{N}} |\psi_i \overline{\psi}_j\rangle \, , \\
& |\psi \overline{\psi}\rangle_\mathbf{\text{Adj}}^A 
= T^A_{ij} |\psi_i \overline{\psi}_j\rangle \, ,
\end{align}
where $T^A$, with $A = 1, \ldots, N^2-1$, are $SU(N)$ generators (in the normalization $\Tr T^A T^B = \delta^{AB}$)
and we properly normalized the states to unitary norm. 
The scattering amplitude in \eq{Mtripletop} has both $SU(3)_C$ and $SU(2)_L$ components. 
In the former case the $\mathcal{S}$-matrix elements in the (color) singlet and adjoint channels are
\begin{align} 
\langle L \bar L | \mathcal{S} |Q \bar Q \rangle_\mathbf{1} &= \frac{\delta_{\alpha\beta}}{\sqrt{3}} \langle L \bar L | \mathcal{S} |Q_\alpha \bar Q_b \rangle 
 = \frac{\delta_{\alpha\beta}}{\sqrt{3}} \mathcal{M}_{SU(3)_C} \delta_{\alpha\beta} = \sqrt{3} \mathcal{M}_{SU(3)_C} \, , \\
\langle L \bar L | \mathcal{S} |Q \bar Q \rangle_\mathbf{\text{Adj}}^A &= T^A_{\alpha\beta} \langle L \bar L | \mathcal{S} |Q_\alpha \bar Q_\beta \rangle 
 = T^A_{\alpha\beta} \, \mathcal{M}_{SU(3)_C} \delta_{\alpha\beta} = 0 \, , 
\end{align}
where $\mathcal{M}_{SU(3)_C}$ denotes the matrix element in \eq{Mtripletop} stripped from the color structure. 
For the $SU(2)_L$ case istead let us collectively denote the doublets (either $Q$ or $L$) by $\psi_a$, with $a=1,2$ being an $SU(2)_L$ index.  
Then the singlet and adjoint scattering channels are
\begin{align} 
_\mathbf{1} \langle \psi \bar \psi  |\mathcal{S}|\psi \bar \psi \rangle_\mathbf{1} 
&= \frac{\delta_{ab}\delta_{cd}}{2} \langle \psi_a \bar \psi_b  |\mathcal{S}|\psi_c \bar \psi_d \rangle
= \frac{\delta_{ab}\delta_{cd}}{2} \mathcal{M}_{SU(2)_L} \left(2 \delta_{ad} \delta_{cb} - \delta_{ab} \delta_{cd}\right) \nonumber \\
&= \frac{1}{2} \mathcal{M}_{SU(2)_L} \left(2 \delta_{aa} - \delta_{aa} \delta_{cc}\right) = 0 \, , \\
_\mathbf{\text{Adj}}^A \langle \psi \bar \psi  |\mathcal{S}|\psi \bar \psi \rangle_\mathbf{\text{Adj}}^B
&= T^A_{ab} T^B_{cd} \langle \psi_a \bar \psi_b  |\mathcal{S}|\psi_c \bar \psi_d \rangle
= T^A_{ab} T^B_{cd} \mathcal{M}_{SU(2)_L} \left(2 \delta_{ad} \delta_{cb} - \delta_{ab} \delta_{cd}\right) \nonumber \\
&= \mathcal{M}_{SU(2)_L} \left(2 \Tr (T^A T^B) - \Tr (T^A) \Tr (T^B) \right) = 2 \delta^{AB} \mathcal{M}_{SU(2)_L} \, , 
\end{align} 
where $\mathcal{M}_{SU(2)_C}$ denotes the matrix element in \eq{Mtripletop} stripped from the $SU(2)_L$ structure.  
Hence, by considering the singlet channel in color space and the adjoint channel 
in $SU(2)_L$ space, we gain respectively a factor $\sqrt{3}$ and $2$ in the partial wave.

\bibliographystyle{utphys.bst}
\bibliography{bibliography}

\end{document}